\documentclass[coloroff]{IEEE_lsens}
\usepackage{textcomp}
\usepackage{graphicx}
\usepackage{siunitx}

\usepackage[noadjust]{cite}
\ifCLASSINFOpdf
\else
\fi
\usepackage[T1]{fontenc} % optional enhanced font encoding
\usepackage{amsmath}
\usepackage[cmintegrals]{newtxmath}
\usepackage{bm}
\usepackage{array}
\usepackage{url}
\ifCLASSINFOpdf
\else
\fi
\providecommand{\hypersetup}[1]{\relax}

\hypersetup{pdftitle={Bare Demo of IEEE\_lsens.cls for IEEE Sensors Letters},%<!CHANGE!
pdfsubject={Typesetting},%<!CHANGE!
pdfauthor={Michael D. Shell},%<!CHANGE!
pdfkeywords={Class, IEEE, IEEE\_lsens, IEEE Sensors Letters, LaTeX, Typesetting, TeX}}%<^!CHANGE!}
\begin{document}

% \markboth{Vol.~1, No.~3, May~2023}{0000000}

% article subject line 
\IEEELSENSarticlesubject{Sensor Applications}

\title{AlGaN/GaN Hall-Effect Sensor for In-Situ Magnetic Field Monitoring of the HSX Stellarator}

\author{\IEEEauthorblockN{Yiming~Zhao\IEEEauthorrefmark{1}, Wayne~Goodman\IEEEauthorrefmark{2}\IEEEauthorieeemembermark{1}, Thomas~Gallenberger\IEEEauthorrefmark{2}, Jasmine~M.~Cox\IEEEauthorrefmark{1}, Benedikt~Geiger\IEEEauthorrefmark{2}, and~Debbie~G.~Senesky\IEEEauthorrefmark{1,3}\IEEEauthorieeemembermark{2}}% <-this % stops a space
\IEEEauthorblockA{\IEEEauthorrefmark{1}Department of Electrical Engineering, Stanford University, Stanford, CA 94305 USA\\
\IEEEauthorrefmark{2}Department of Nuclear Engineering and Engineering Physics, University of Wisconsin-Madison, Madison, WI 53706 USA\\
\IEEEauthorrefmark{3}Department of Aeronautics and Astronautics, Stanford University, Stanford, CA 94305 USA\\
\IEEEauthorieeemembermark{1}Member, IEEE\\
\IEEEauthorieeemembermark{2}Senior Member, IEEE}%
% LSENS authors should provide a real e-mail address here.
\thanks{Corresponding authors: Yiming Zhao (e-mail: timzhao@stanford.edu), Debbie G. Senesky (e-mail: dsenesky@stanford.edu).}}% <-this % stops a space
\IEEEtitleabstractindextext{%
\begin{abstract}[graphical_abstract]
Direct magnetic field sensors can address integration drift commonly observed in conventional inductive magnetic diagnostics used in fusion systems. In this work, an AlGaN/GaN Hall-effect sensor was fabricated, packaged, and deployed inside the Helically Symmetric eXperiment (HSX)---the first quasi-helically symmetric stellarator, operating with a 1 T on-axis magnetic field and up to 200 kW of launched electron cyclotron resonance heating (ECRH) power---for in-situ magnetic field monitoring near the plasma edge. The sensor leverages the high-mobility two-dimensional electron gas (2DEG) formed in the AlGaN/GaN heterostructure for sensitive magnetic field measurement, while the wide-bandgap GaN material system provides thermal robustness for harsh-environment operation. During 68 consecutive plasma discharge shots, the sensor remained functional and produced clear transient responses associated with plasma ignition and discharge dynamics. Comparisons between biased and unbiased operation, as well as plasma-discharge and coil-only shots, confirmed that the response originated from the biased Hall-effect sensor element. Furthermore, the sensor output exhibited temporal correlation with the plasma stored energy measured by the HSX diamagnetic loop across high-energy, late-breakdown, and failed-breakdown discharges.
\end{abstract}

\begin{IEEEkeywords}
Magnetic sensors, Hall-effect sensors, gallium nitride, plasma diagnostics, stellarators.
\end{IEEEkeywords}}

% \IEEEpubid{}  % copyright/pubid line removed

\maketitle

\section{Introduction}
\IEEEPARstart{M}{agnetic} confinement fusion offers the prospect of clean, carbon-free baseload energy, and is regarded as a critical long-term option for future energy supply~\cite{Ongena2016}. Modern fusion experiments increasingly rely on data-driven modeling and machine-learning--based control, placing growing demands on in-situ plasma diagnostics and the associated instrumentation~\cite{Degrave2022, Anirudh2023}. Magnetic field diagnostics are essential for monitoring plasma position, shape, stability, and stored energy in magnetic confinement fusion devices such as tokamaks and stellarators~\cite{ref1}. As fusion experiments advance toward long-pulse and steady-state operation~\cite{ref2}, in-situ diagnostics must provide reliable, real-time measurements of plasma-generated magnetic fields for equilibrium reconstruction and control. Conventional diagnostic systems with inductive sensors, such as Mirnov coils, recover the magnetic field by integrating the induced voltage proportional to its time derivative. During extended operation, this integration becomes problematic as it accumulates low-frequency noise and radiation-induced offsets, leading to drift and reduced accuracy in tracking magnetic fields~\cite{ref3}, \cite{ref4}.

Hall-effect sensors offer an attractive alternative because they enable direct magnetic field measurement without integration, eliminating drift at the source~\cite{ref5}, \cite{ref6}. However, conventional Hall-effect devices based on silicon, gallium arsenide (GaAs), indium arsenide (InAs), and indium antimonide (InSb) platforms cannot be deployed near the plasma edge because the in-vessel environment combines elevated temperature, vacuum operation, chemical reactivity, and radiation exposure. Gallium nitride (GaN) is a wide-bandgap semiconductor with strong thermal tolerance, making it well suited for harsh-environment electronics and in-situ sensing~\cite{ref7}, \cite{ref8}. In AlGaN/GaN heterostructures, a two-dimensional electron gas (2DEG) forms at the interface~\cite{ref9}. Consequently, the high electron mobility of this 2DEG channel enables high sensitivity ideal for this application~\cite{ref10}. In prior thermal characterization experiments, AlGaN/GaN Hall-effect sensors maintained stable current-scaled sensitivity from room temperature up to \SI{576}{\degreeCelsius}~\cite{ref11}, supporting their potential for extreme-environment magnetic field sensing~\cite{ref12}.

In this work, we demonstrate an AlGaN/GaN Hall-effect sensor for in-situ magnetic field monitoring inside the Helically Symmetric eXperiment (HSX), the first stellarator designed and built with a quasi-helically symmetric magnetic field, operated at a \SI{1}{\tesla} on-axis magnetic field~\cite{ref13}, \cite{ref14}. The Hall-effect sensor was fabricated, packaged for ultra-high-vacuum and high-temperature operation, and deployed near the plasma edge. By comparing biased and unbiased operation, plasma-discharge and coil-only shots, and the sensor output with conventional plasma stored-energy diagnostics, we validate that the packaged AlGaN/GaN Hall-effect sensor can detect plasma-dependent magnetic field dynamics in real time, with a 1 MHz readout bandwidth across 68 consecutive shots.
\newpage
\section{Sensor Fabrication and Experimental Setup}
\subsection{Fabrication}
\begin{figure}[!t]
\centering
\includegraphics[width=\columnwidth]{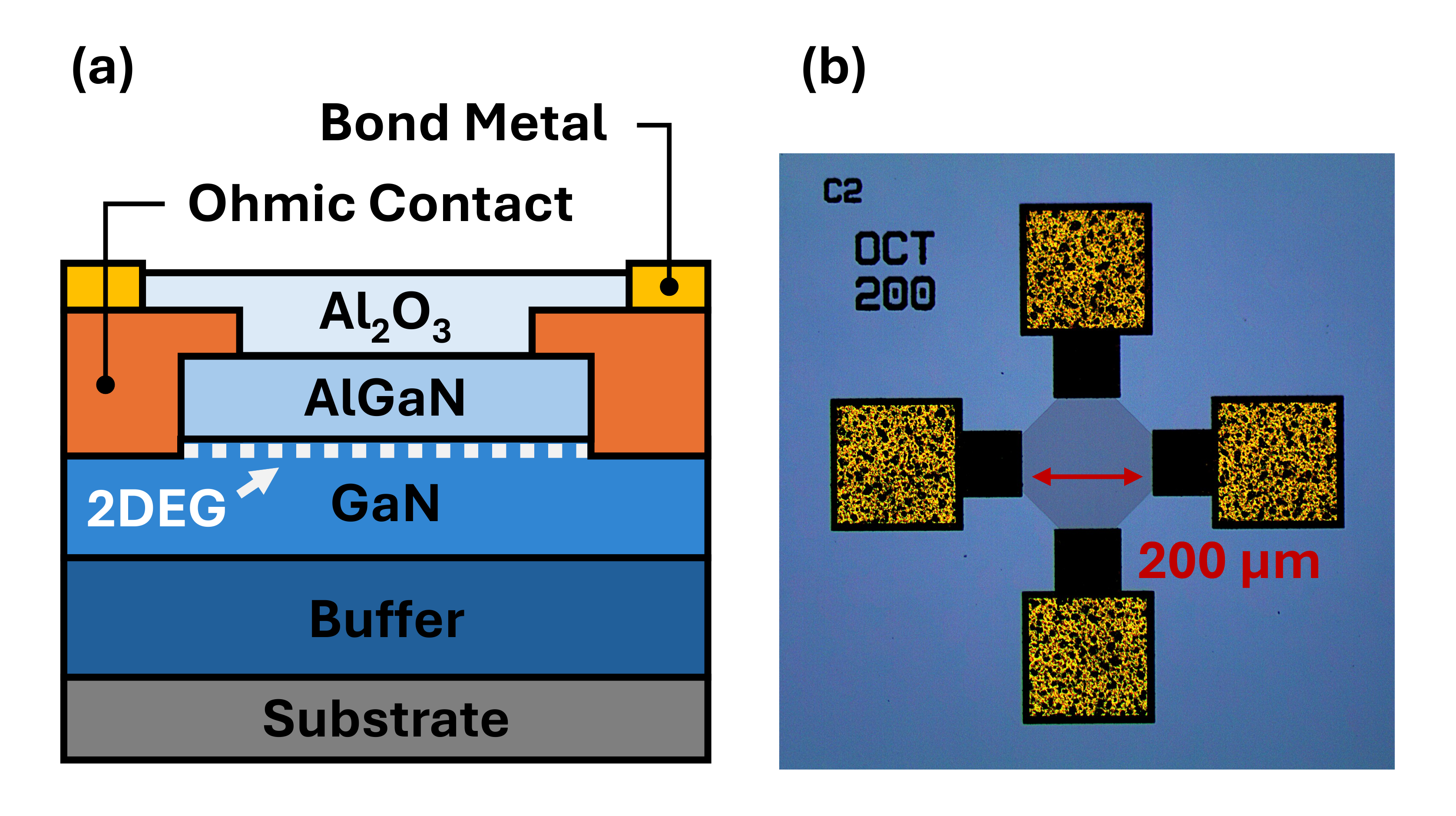}
\caption{Fabricated AlGaN/GaN Hall-effect sensor. (a) Cross-sectional schematic of the passivated AlGaN/GaN Hall-effect sensor showing the 2DEG sensing layer, Ohmic contacts, Al$_2$O$_3$ passivation, and bond metal. (b) Optical image of the fabricated regular octagonal Hall-effect plate with a 200 \textmu m inscribed diameter.}
\label{fig:fab}
\end{figure}

The AlGaN/GaN wafer, purchased from NTT Advanced Technology Corporation, consists of a \SI{3.7}{\micro\meter} buffer structure, a \SI{300}{\nano\meter} GaN layer, a \SI{1}{\nano\meter} AlN spacer, and a \SI{22}{\nano\meter} Al$_{0.28}$Ga$_{0.72}$N barrier layer. The Hall-effect sensors were fabricated at the Stanford Nanofabrication Facility. The fabrication process began with a mesa etch of the III-nitride layer to define the active regions, followed by deposition and rapid thermal annealing of a Ti~(\SI{25}{\nano\meter})/Al~(\SI{200}{\nano\meter})/Mo~(\SI{40}{\nano\meter})/Au~(\SI{80}{\nano\meter}) stack at \SI{850}{\degreeCelsius} for \SI{35}{\second} to form Ohmic contacts~\cite{ref11}. A \SI{7}{\nano\meter}-thick Al$_2$O$_3$ passivation layer was then deposited; vias were etched to expose the contacts; and a Ti/Au bond-metal layer was deposited for wire bonding. The wafer was singulated into \SI{5}{\milli\meter} $\times$ \SI{5}{\milli\meter} dies for packaging. Fig.~\ref{fig:fab}(a) shows a cross-sectional schematic of the AlGaN/GaN Hall-effect sensor. The fabricated Hall-effect sensors feature a regular octagonal geometry with an inscribed diameter of \SI{200}{\micro\meter}~\cite{ref10}, as illustrated in Fig.~\ref{fig:fab}(b).
\subsection{Packaging}
\begin{figure}[!t]
\centering
\includegraphics[width=\columnwidth]{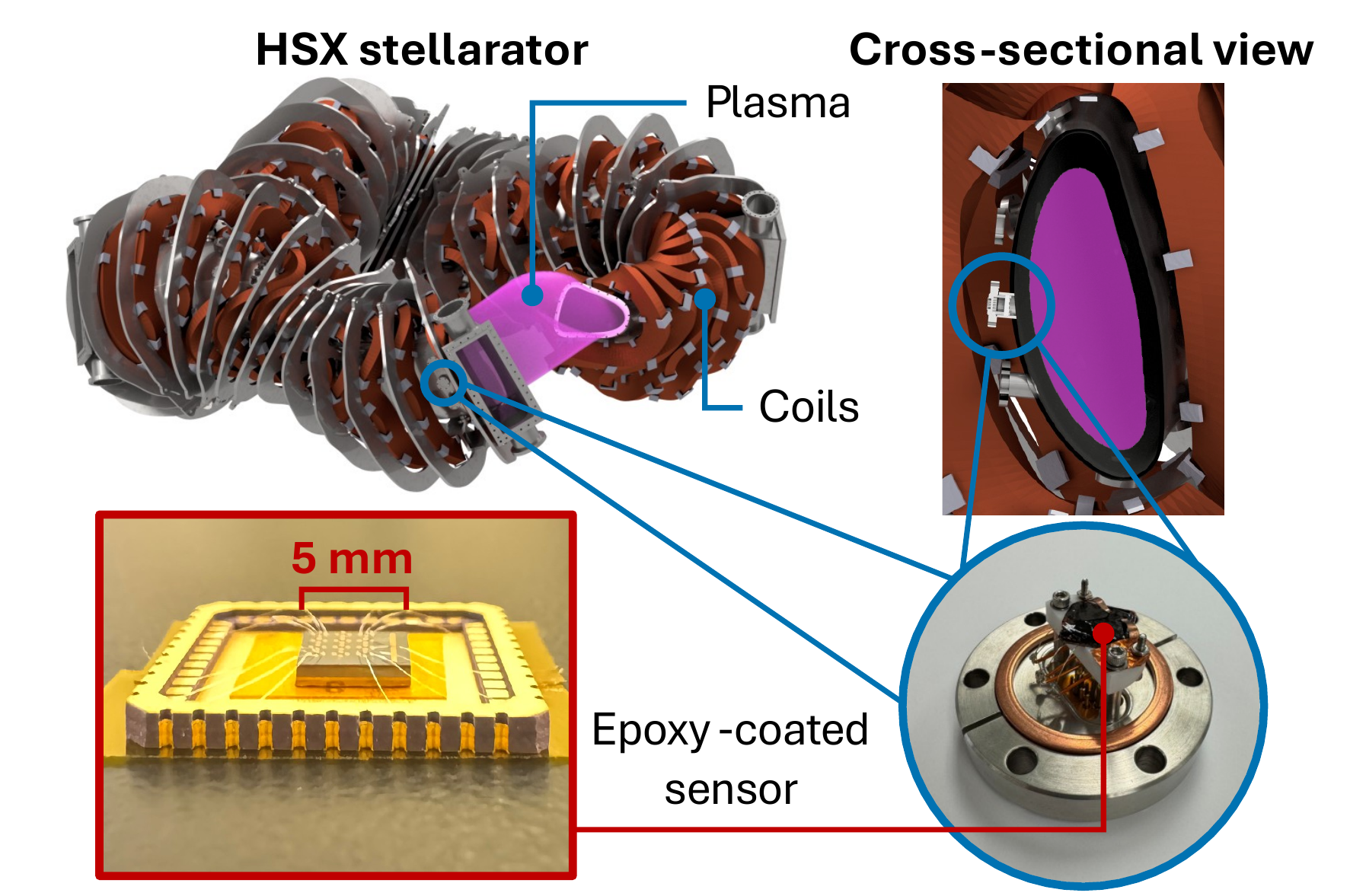}
\caption{Packaging and in-vessel deployment of the AlGaN/GaN Hall-effect sensor in HSX. The sensor die was wire-bonded to a ceramic leadless chip carrier, epoxy-coated, and mounted on a custom flange for insertion into the HSX vessel near the plasma edge.}
\label{fig:pkg}
\end{figure}

The individual die was wire-bonded to a ceramic leadless chip carrier (LCC; Spectrum Semiconductor Materials) using aluminum wire. The die and wire bonds were then encapsulated with epoxy (EPO-TEK 353ND) and vacuum-baked at \SI{150}{\degreeCelsius} for 1~hour to meet the ultra-high-vacuum (UHV) requirements of the HSX facility. The packaged sensor was mounted on a custom zirconia ceramic holder attached to a stainless-steel standoff for insertion into the HSX vessel. To reduce the risk of arcing and epoxy degradation during glow discharge cleaning (GDC) and plasma operations, a grounded graphite shield was installed over the packaged sensor module. The packaged sensor assembly and its in-vessel placement are shown in Fig.~\ref{fig:pkg}.
\begin{figure}
\centering
\includegraphics[width=\columnwidth]{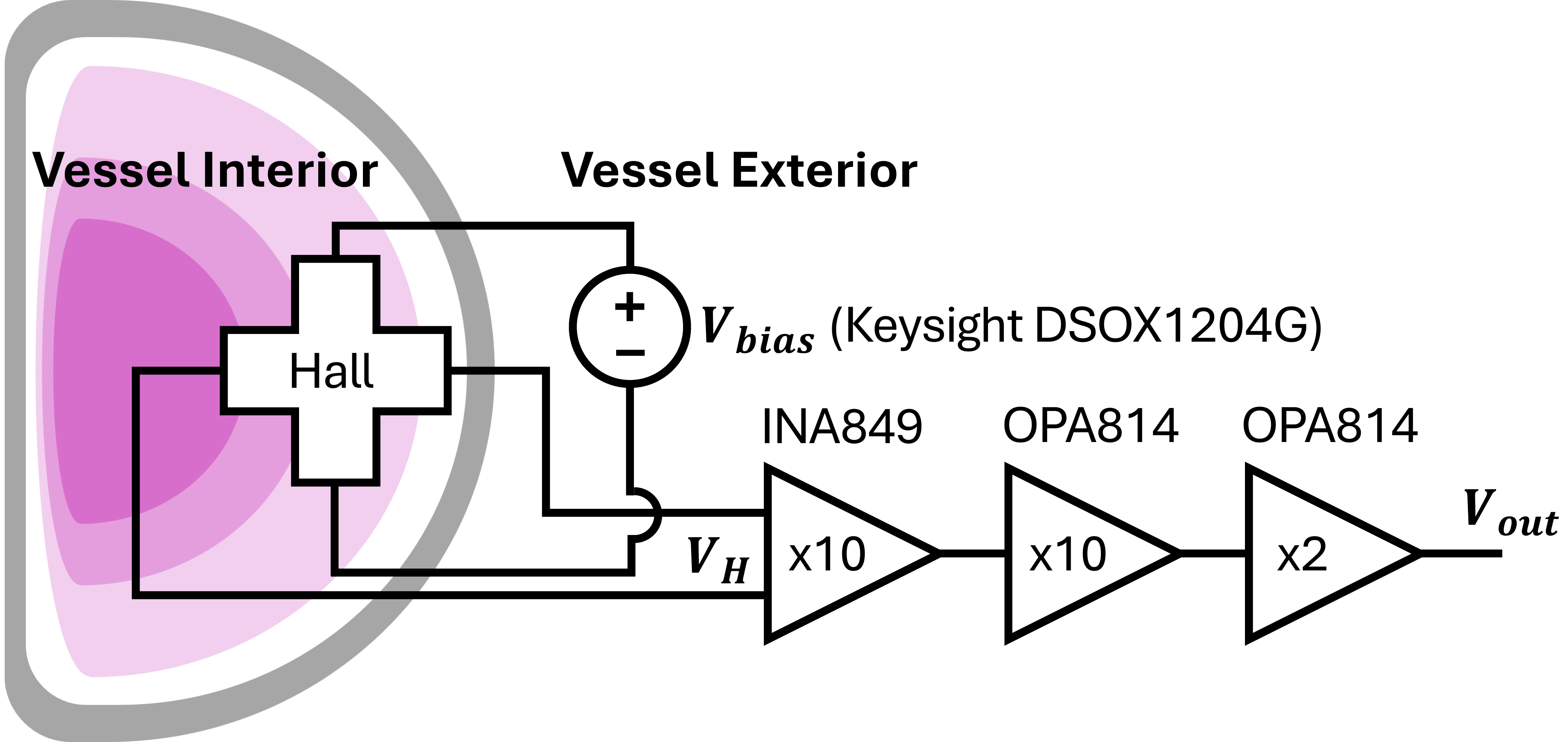}
\caption{Experimental readout configuration for the in-vessel Hall-effect sensor. The sensor was voltage-biased using the built-in waveform generator of an oscilloscope, and the transverse Hall voltage was amplified using a differential readout chain before data acquisition.}
\label{fig:readout}
\end{figure}
\subsection{Experimental Setup}
The Hall-effect sensor module was deployed inside the HSX vessel near the plasma edge, as shown in Fig.~\ref{fig:pkg}. The bias supply, readout electronics, and data acquisition system were located outside the vessel and connected to the sensor through vacuum feedthroughs. During operation, the sensor was voltage-biased and generated a transverse Hall-effect voltage given by
\begin{equation}
V_{H} = S_{v} V_{\mathrm{bias}} B,
\end{equation}
where $S_{v}$ is the voltage-scaled sensitivity, $B$ is the orthogonal magnetic field, and $V_{\mathrm{bias}}$ is the bias voltage. As illustrated in Fig.~\ref{fig:readout}, two opposite terminals of the Hall-effect plate were biased using the built-in waveform generator of an oscilloscope (Keysight DSOX1204G), while the Hall-effect voltage was measured across the orthogonal terminals. The differential Hall-effect voltage was amplified by a readout circuit consisting of an instrumentation amplifier (Texas Instruments INA849) followed by two operational amplifier (Texas Instruments OPA814) stages, providing a total voltage gain of \SI[per-mode=symbol]{200}{\volt\per\volt} and a bandwidth of \SI{1}{\mega\hertz}. The amplified output voltage can be expressed as
\begin{equation}
V_{\mathrm{out}} = A_{v} V_{H} + V_{\mathrm{off}},
\end{equation}
where $A_{v}$ is the voltage gain and $V_{\mathrm{off}}$ accounts for offset contributions from the sensor and amplifier stages. The amplified output was continuously monitored during the plasma discharge shots. In this readout chain, $V_{\mathrm{off}}$ is bias-independent but may vary with temperature during operation~\cite{ref15}. Since $V_{H}$ scales linearly with $V_{\mathrm{bias}}$ whereas $V_{\mathrm{off}}$ does not, comparing the sensor output under biased ($V_{\mathrm{bias}} = \SI{0.4}{\volt}$) and unbiased ($V_{\mathrm{bias}} = \SI{0}{\volt}$) operation isolates the bias-dependent Hall-effect response from the offset and from other artifacts such as electromagnetic pickup or charge-induced response. Absolute calibration of $V_{\mathrm{off}}$ and quantitative correction for its temperature dependence remain future work.
\begin{figure}[!t]
\centering
\includegraphics[width=\columnwidth]{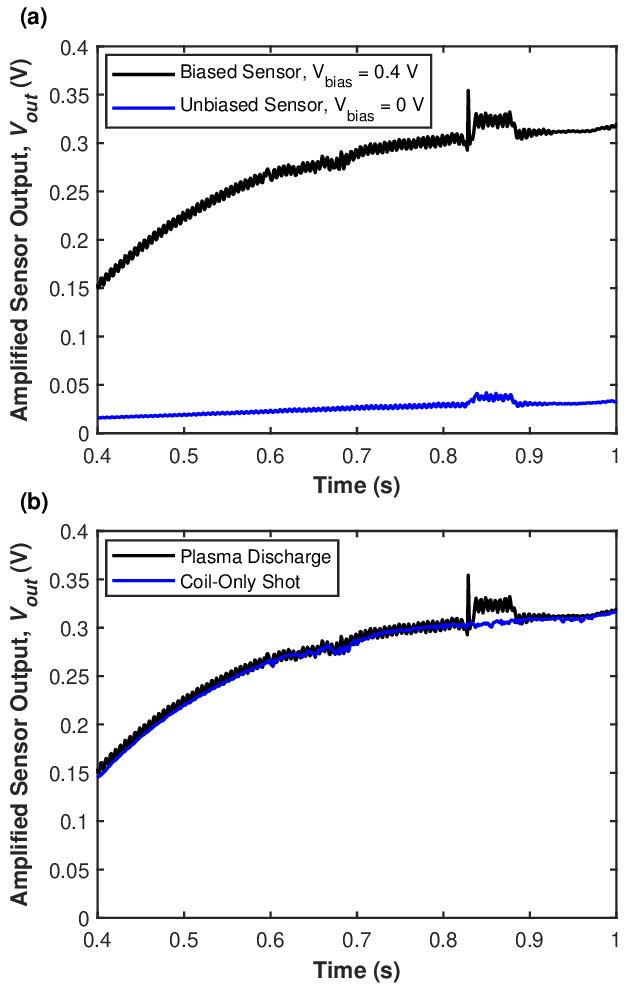}
\caption{Functional verification of the packaged AlGaN/GaN Hall-effect sensor during HSX operation. (a) Amplified sensor output for unbiased (shot 63) and biased (shot 65) operation. (b) Comparison of the biased sensor response during a plasma discharge (shot 65) and a coil-only shot (shot 68).}
\label{fig:func}
\end{figure}
\section{Results and Discussion}
\subsection{Sensor Functionality}
The packaged AlGaN/GaN Hall-effect sensor was deployed inside the HSX stellarator and operated across 68 consecutive shots. Prior to each plasma discharge, mechanical energy stored in 18 motor generators was converted to electrical current driving the 48 magnetic field coils of HSX; after approximately \SI{800}{\milli\second} the coils' current saturated, establishing the target magnetic confinement field. Plasma was then ignited within the resulting \SI{50}{\milli\second} flat-top period using up to \SI{200}{\kilo\watt} of \SI{28}{\giga\hertz} electron cyclotron resonance heating (ECRH). Fig.~\ref{fig:func}(a) compares the amplified output of the Hall-effect sensor when unbiased ($V_{\mathrm{bias}} = \SI{0}{\volt}$; shot 63) and when biased at $V_{\mathrm{bias}} = \SI{0.4}{\volt}$ (shot 65). Consistent with the Hall-effect voltage dependence on $V_{\mathrm{bias}}$, the unbiased output remains near a low baseline, whereas the biased output increases during coil ramp-up and shows a sharp transient followed by fluctuations during the plasma discharge. To separate the plasma-induced response from the background coil field, Fig.~\ref{fig:func}(b) compares a plasma discharge (shot 65) with a coil-only shot (shot 68). Both traces show a similar increase during coil ramp-up, but only the plasma discharge exhibits the pronounced transient and subsequent fluctuations near plasma ignition. These results confirm that the packaged sensor remained functional in the HSX environment and detected magnetic field variations associated with plasma dynamics, with the ignition transient consistent with the diamagnetic response of the confined plasma as plasma pressure rises and expels magnetic flux. Across the 68 discharge shots, the shape and timing of this transient repeated reliably under comparable conditions, indicating stable sensor operation throughout the deployment.
\begin{figure}[!t]
\centering
\includegraphics[width=\columnwidth]{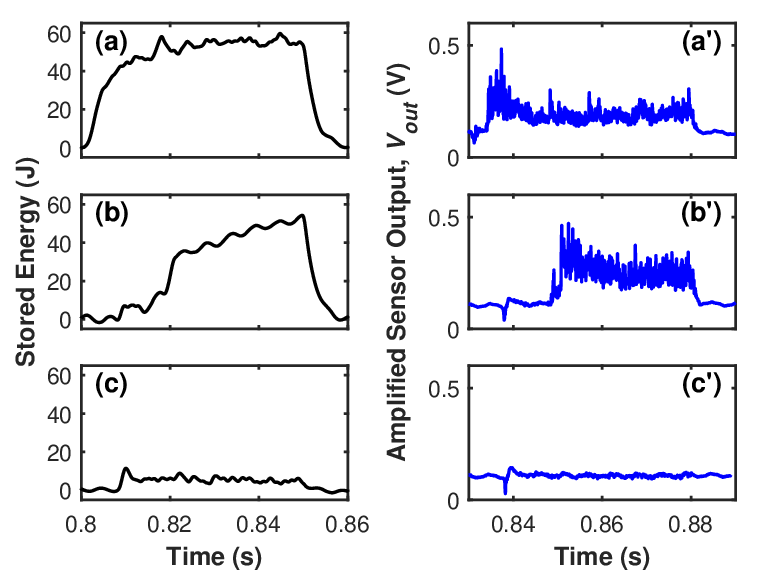}
\caption{Real-time tracking of plasma-dependent magnetic-field dynamics. Plasma stored energy from conventional HSX diagnostics is compared with the amplified AlGaN/GaN Hall-effect sensor output for (a), (a$'$)~a high-energy discharge (shot 21); (b), (b$'$)~a late-breakdown discharge (shot 18); and (c), (c$'$)~a failed-breakdown case (shot 19). The two data acquisition systems have a $\sim$30 ms relative timing offset due to delayed triggering.}
\label{fig:track}
\end{figure}
\subsection{Real-Time Plasma Energy Tracking}
The measurements shown in Fig.~\ref{fig:track} compare the amplified output of the in-vessel AlGaN/GaN Hall-effect sensor with the plasma stored energy measured by the HSX diamagnetic loop diagnostic~\cite{ref16} for three representative discharge conditions. The two diagnostic methods measure physically distinct quantities: the Hall-effect sensor probes the local magnetic field at a single in-vessel location, whereas the diamagnetic loop measures the change in enclosed toroidal flux from which volumetric plasma stored energy is inferred. Both, however, respond to plasma magnetic dynamics and are therefore expected to be temporally correlated~\cite{ref1}. The two signals are acquired through separate data acquisition systems with a known $\sim$\SI{30}{\milli\second} relative timing offset arising from delayed triggering of the diamagnetic loop DAQ. In the high-energy discharge (shot 21) shown in Fig.~\ref{fig:track}(a) and (a$'$), the rapid rise and sustained stored-energy plateau are accompanied by a pronounced sensor response and elevated fluctuations during the discharge. In the late-breakdown case (shot 18) shown in Fig.~\ref{fig:track}(b) and (b$'$), the sensor output remains near baseline until the delayed rise in plasma stored energy, after which a clear increase in signal amplitude is observed. In contrast, for the failed-breakdown case (shot 19) shown in Fig.~\ref{fig:track}(c) and (c$'$), both the stored energy and the sensor output remain near baseline. Although quantitative calibration remains future work, the consistent temporal agreement across these discharge scenarios establishes that the AlGaN/GaN Hall-effect sensor can track plasma-dependent magnetic-field dynamics in real time.
\section{Conclusion}
An AlGaN/GaN Hall-effect sensor was fabricated, packaged, and deployed inside the HSX stellarator (\SI{1}{\tesla} on-axis field, up to \SI{200}{\kilo\watt} of ECRH) for in-situ magnetic-field monitoring near the plasma edge. The packaged device operated through 68 consecutive shots under ultra-high-vacuum and elevated-temperature conditions. Comparisons between biased and unbiased operation, together with plasma-discharge and coil-only measurements, confirmed that the observed response was associated with magnetic field changes detected through the biased Hall-effect sensor. The amplified sensor output further showed temporal agreement with conventional plasma stored-energy diagnostics across high-energy, late-breakdown, and failed-breakdown scenarios. Although quantitative calibration remains future work, these results establish AlGaN/GaN Hall-effect sensors as a feasible platform for real-time magnetic diagnostics in fusion-relevant environments. Future work will focus on absolute calibration through in-situ cross-reference with established HSX magnetic diagnostics~\cite{ref17}, extended-duration deployments in HSX to evaluate offset stability, radiation and neutron irradiation characterization at a dedicated facility to assess radiation hardness, and integration of lower-noise readout electronics for resolving smaller-amplitude magnetohydrodynamic fluctuations.


% Appendixes, if needed, appear before the acknowledgment.
\section*{Acknowledgment}
\addcontentsline{toc}{section}{Acknowledgment}
\scriptsize
The work of the authors was supported by the U.S.\ Department of Energy under Contract No.~DE-AC02-76SF00515, SLAC FWP~101264, and by the TomKat Center for Sustainable Energy at Stanford University. Fabrication work was performed at the Stanford Nanofabrication Facility, a member of the National Nanotechnology Coordinated Infrastructure (NNCI), supported by the National Science Foundation under Award~ECCS-2026822.
% put at least one blank line to end the scriptsize paragraph and
% then revert back to normalsize.
\normalsize

% Last page column equalization
%
% IEEE Sensors Letters does balance the columns on the last page.
% Can use:
% \IEEEtriggeratref{8}
% to trigger a \newpage just before the given reference number to
% balance the columns on the last page. Adjust the reference number
% as needed - this may need to be readjusted if the document is 
% modified later.
% Alternatively, you can also directly use something like
% \enlargethispage{-7in}
% on the last page instead of breaking at a specific reference number.

%\bibliographystyle{IEEEtran}
% Before submitting to IEEE Sensors Letters, manually copy in the
% resultant .bbl file contents in place of the \bibliographystyle and
% \bibliography lines here:

\end{document}